# The impact of Over The Top service providers on the Global Mobile Telecom Industry: A quantified analysis and recommendations for recovery

**Ahmed Awwad**
*Faculty of Science and Engineering, Anglia Ruskin University, Cambridge Campus, East Rd, Cambridge CB1 1PT, United Kingdom.*
ahmed_saied_02@hotmail.com

**Key words:** Mobile Network Operators (MNO), Over the top (OTT), Telecom industry revenue decrease, telecom operator's revenue growth strategies.

# ABSTRACT

Telecom industry is significantly evolving all over the globe than ever. Mobile users' number is increasing remarkably. Telecom operators are investing to get more users connected and to improve user experience, however, they are facing various challenges. Decrease of main revenue streams of voice calls, SMS (Short Message Service) and LDC (Long distance calls) with a significant increase in data traffic. In contrary, with free cost, OTT (Over the top) providers such as WhatsApp and Facebook communication services rendered over networks that built and owned by MNOs. Recently, OTT services gradually substituting the traditional MNOs` services and became ubiquitous with the help of the underlying data services provided by MNOs. The OTTs` services massive penetration into telecom industry is driving the MNOs to reconsider their strategies and revenue sources.

**Objective** – The research objective is to critically evaluate the impact of OTT providers on MNOs. With the empirical evidences, the research explores a quantified impact of OTTs on MNOs main revenue streams, and breakdown this impact to highlight effect`s significance. Followed by a statistical trend analysis of global MNOs` revenue over the last decade. Additionally, the research aims to compile, evaluate, and analyse set of former proposed strategies for MNOs to overcome the OTTs` impact.

**Methods** – Referring to available open source raw data collected from GSMA and official telecommunications regulatory authorities worldwide, the research develops a statistical model based on regression and extrapolation to analyse the global MNOs` revenue trend.

**Findings** – This study reveals a hidden revenue loss for global telecommunications industry represented in all MNOs all over the world. The study corelates the hidden revenue loss to OTTs evolution. The study finds out that SMS is the most impacted traditional service of MNOs and predicted that it will vanish at quarter 4 of 2022.

**Recommendations** – MNOs to apply 4 different strategies varying from competing to partnering with OTTs based on the services overlapping level and OTT size. MNOs need to target OPEX and CAPEX reductions and develop their own multi-functional customer-oriented OTT application.

# I   Introduction

## I.I   Background

Last decade has witnessed a revolutionary development of global mobile telecommunications industry while playing the crucial role in empowering people, triggering positive change in business processes and in the global economy with 4.1 $ trillion contribution in GDP (Gross Domestic Product) at 2019 as per GSMA (2020a). Mobile users' numbers are increasing significantly day by day and generating tremendous increase in network's load and traffic. Pursuant to GSMA (2020b), the unique number of mobile subscribers has reached 5.2 billion at April 2020, meanwhile Worldometers (2020) reported 7.77 billion world population during the same month, that means out of 7.77 , 5.2 are mobile users, with a users` penetration rate reaches 66.9 % of mobile telecom industry represented in MNOs (Mobile Network Operators) around the world, notwithstanding, the MNOs` revenues are declining.

The barriers to entry to telecommunications market have been lowered due to the widespread adoption of mobile internet access (Fritz et al. 2011), specifically with the 4G (4th Generation) mobile technology advancements in telecom industry, OTTs (Over The Top) have entered the market (Legere 2018). OTTs refers to the mobile data services that carried over the MNOs` networks (Minges and Kelly 2018). Those OTTs make use of MNOs' infrastructure and internet capacity rather than creating their own networks (Valipour and Hosseini 2017). Furthermore, OTTs have succeeded to shift the users from using MNOs` services towards OTTs (Esmailzadeh 2016). Whilst incumbent MNOs battling to churn those users back, OTT providers steadily keep growing their customer bases and revenues (Infopulse 2019).

The user`s shift to OTTs caused a decline in traditional services offered by MNOs including domestic voice calls, text messages and long-distance calls, accompanied with a disruptive data traffic increase (Minges and Kelly 2018) and (Cisco 2018). That resulted into a declining trend of MNO's traditional revenue streams (SMS and voice) (Cisco 2018), while OTTs are developing with an extensive increase in revenues (Ouyang et al. 2018) and (Sujata et al. 2015). On the other hand, this enormous increase in data traffic is forcing the MNOs` investment directions towards expanding their network`s capacity to accommodate the increase in traffic while maintaining good level of customer experience (Czarnecki and Dietze 2017). In this challenging situation, MNOs have to revise their strategies to keep competing with OTTs, as these OTT service providers increasingly substituting traditional telecommunications services due to their alternative offered services (Trask 2018).

## I.II   Problem statement

The 66.9 % penetration rate of telecommunications industry reflects its value and urge to explore its major opportunities and potential threats in order to keep this industry flourishing. Mobile telecommunications industry is confronting several challenges in this disruptive decade. Immense surge in mobile networks` data traffic due to the proliferation of smartphones penetration and the emergence of data-based applications (Cisco 2018), which necessitates the MNOs` networks expansion in order to sustain the quality of the increased traffic (Czarnecki and Dietze 2017). In spite of the massive subscribers' numbers increase and data outgrowth, the revenue growth rates and the ARPU (Average Revenue Per User) MNOs` trends are declining (GSMA 2020b), which means, MNOs are not able to monetize their investments. Meffert and Mohr (2017) predicted 15 to 30 % revenue loss, if MNOs will only expand and modernize their network without additional recovery strategies.

The content and services offered by OTT providers in today`s world like Facebook, Whatsapp, Spotify and Netflix became part of every day's life of billions of people all over the world, those services rapidly acknowledged by tech-savvy millennials who were born mainly from 1980`s to 2000`s (Krämer and Jalajel 2019). This whopping rise of gargantuan OTTs have absorbed a large fraction of MNOs` services traffic (Krstevski and Mancheski 2016) and might have resulted into massive reduction in MNOs` revenues (Sujata et al. 2015). The traditional voice and messaging traffic of MNOs have been replaced by data traffic, which is generated from using the OTT applications in which customer can access premium functionalities with a subscription fees to OTTs however the basic features are with no cost. Users are still paying for data consumption usage to the MNOs but OTTs using freemium model at which they utilize the MNOs networks without paying to them, so they generate high revenues (Minges and Kelly 2019). Core services of MNOs are shrinking, as a result of the traditional transport and access business becomes negative, which may fail to achieve the broadband goals (Knoben 2015). In 2017 MNOs have generated 1.3 trillion Euro compared to 402 billion Euro from OTTs, however the OTT growth by a CAGR (Compound annual growth rate) 2012-2017 of 17.3%, MNOs only 2.6% (Knoben 2015).

MNOs` ARPU is driven down with the rise of OTT services` popularity. As shown at Figure 1 (amended from GSMA 2020b), the mobile user's number is enormously growing, unlikely, the ARPU is decreasing, which means MNOs are not able to monetize the increased numbers in subscribers. Even though the subscriber's number increased by 83% comparing 2009 to 2019, the revenue has only increased by 32.67% and ARPU has fallen down to $ 13.56 at 2019 compared to $ 22.39 at 2009. Knoben (2015) warned that revenues from core MNOs access and transport services are declining, and urged MNOs to look for new sources of income while the OTTs are stepping inside the value chain and causing a fierce competition, the MNOs current investments in ultra-fat reliable infrastructure may fail to achieve the policy goals. This protrude an earnest issue happening with the MNOs. MNOs are rethinking their business models and attempting to reinsert back the traditional telecom services into the current telecommunications equation (Athow 2015). This challenge engenders a substantial pressure on MNOs to recoup the revenue loss if it is relevant to OTTs, and to review their stagnant strategies in order to monetize their ongoing investments (EIU 2018).

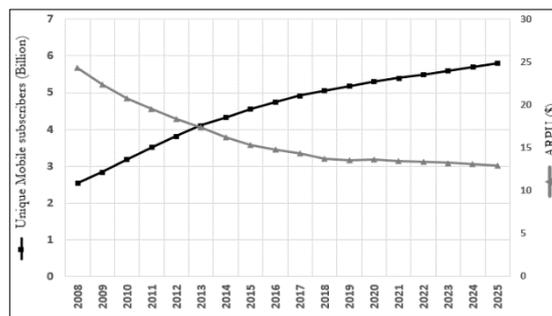

Figure 1. MNOs subscriber's vs APRU trend (amended from GSMA 2020b)

Recapping the problem; the mobile users' numbers are significantly increasing. Despite this increase, the revenue growth rate and ARPU are falling. Which means MNOs are not able to monetize their investments that includes infrastructure expansion to accommodate the traffic increase generated from the increased mobile users' numbers. Disability of MNOs to monetize their investments will results in deteriorating the industry and leave the 2.5 Billion unconnected people in the world offline. The emergence of OTTs can be a major factor in impacting the telecommunications industry, OTTs are using MNOs` network infrastructure to deliver services to users that replaces MNO`s services and cannibalizing traditional MNO`s revenue streams. Due to the quick embracing of OTTs in society, the MNOs are being reactive and slow till now to deal with the OTTs impact which soars the problem more and develops innovation backlog of MNOs against OTTs.

## I.III  Statement of purpose

Several researches have attempted to examine the OTTs` impact on MNOs from various perspectives; nevertheless, this impact was not quantified neither corelated. This research takes a cognisance of the impact of OTTs on MNOs from revenue perception. Additionally, the research compares MNOs vs OTTs trends of LDC, Voice calls, SMS and traffic which illustrates the impact on MNOs from OTTs. The research acknowledges the MNOs` data traffic increased with OTTs evolution which has given rise to MNO`s revenues, though not to the same revenue level they used to earn from traditional services as LDC, voice calls and SMS. To analyse these variations of the mobile telecommunications industry revenue along with the introduction of OTTs, the study targets the period from 2009 till 2025 where it is divided into 3 phases as shown at table 1 (personal collection). Phase I is the prosperous era of the industry when it recorded the highest revenue growth rate. Phase II is the recent era which witnessed the evolution of OTTs with a decreasing revenue growth factor of MNOs. Lastly, Phase III is the future era. The research takes advantage of MNOs phase I`s success as a reference baseline to reveal a quantified revenue loss at phase II, furthermore, recommend a two-dimensional strategy for MNOs to contain the OTTs impact and to prevent the loss at phase III.

| Phase I | | Phase II | | Phase III | |
|---------|-----|----------|------|-----------|------|
| From    | To  | From     | To   | From      | To   |
| 2008    | 2013| 2014     | 2019 | 2020      | 2025 |

**Table 1. Research`s period scope (personal collection)**

Revenue loss quantification gives an explicit view of the direct impact occurred by the OTTs, after corelating the loss to them. Also, it urges the regulators and MNOs to push forward on carrying out former proposed researches` strategies to compete with the OTTs, or strategies to regulate the OTTs usage of MNOs. Moreover, the research studies the decreasing ARPU trend of MNOs vs an OTT ARPU, defining the turnover juncture point at which the 2 ARPUs have been equalized and then the MNOs` ARPU started to be less than OTTs` ARPU which kept increasing. Defining this turnover point will help to precisely predict the quantified revenue loss of MNOs and correlates it to the OTT impact. The study combines several researchers` proposals to overcome the OTTs` impact on MNOs. This research combines and analyse 27 former proposed strategies, to result in recommending a combined competitive-participatory approach for MNOs to avoid further revenue losses caused by OTT providers in the future. The derived research`s results in relation to the quantified revenue loss empowers the recommended strategy and pushforward implementing it, with ability to measure the improvement.

Here comes the research logic; with the advancements of telecom industry, mobile users' number is increasing, and MNOs are funding more investments to serve the increased traffic. Unlikely, the ARPU and revenue growth rate are deteriorating. The dispute is about if the OTTs impacted the declining trends. The study recommends a hybrid strategy be followed by MNOs` in order to avoid revenue forfeiture and get their declining revenue growth trends back, where this change is inevitable for MNOs. The research aims to achieve 2 objectives, first is to critically evaluate the impact of OTT service providers on global MNOs. Second is to critically evaluate the strategies proposed for MNOs to overcome the OTTs` impact. In order to get hold of the objectives, the research questions have been designed relatively by applying techniques of SMART project management following Grant et al. (2013), the questions have been derived as below:

[1] Is the emergence of OTTs presents a threat or opportunity to MNOs?
[2] What are the main revenue streams of MNOs, what is the impact brought by OTT providers on those streams? What is the most affected stream? how the MNOs vs OTTs services` trends looks like?
[3] Why customers shifted to OTTs` voice and messaging services?
[4] What is the hidden revenue loss for MNOs, what is the quantified delta of MNO's revenue growth if phase I and phase II compared? How much is the amount of the revenue loss?
[5] What are strategies proposed by former researches to overcome the OTTs` impact on MNOs? What is the most popular recommended strategy in the researches?
[6] What MNOs should follow to improve their revenue streams in phase III?

## I.IV    Potential value of the research

In view of the current robust penetration of telecommunications industry in the world, the conclusions of this research redound to the advantage of humanity and society; flourishing telecommunications industry back will help in connecting 2.5 unconnected people around the world. With a specific affirmation on the impact of OTT on MNO's revenues, this study is envisaged to make a workable participation to the telecom industry body of knowledge. The revenue loss reveal developed model in this research can be further utilized by MNOs enabling them to expose their hidden revenue loss. The research recommendations of how to overcome the OTTs impact on MNOs, shall go for implementation by MNOs quicker than former researchers` proposals, as it is empowered by an actual quantified figure of the OTTs impact. Applying the two-dimensional strategy will convert the revealed unseen revenue loss into an actual revenue for MNOs` that will help to prosper the global mobile telecom industry. In the light of the revealed hidden revenue loss, this study helps to expedite the other researcher's former recommended strategies which have been pending for long time without implementation.

## II  Methodology

While the usual format is for the literature review section to come before the Methodology section (Nikolov 2013), in this research the methodology section has been written first. As the literature review is a method used to achieve 2 objectives of the research. Additionally, the literature review is including some supporting analytics pulling former researches together that will be explained at methodology section.

Methods are the processes and vehicles used in order to collect and analyse the data sets in this research (Wisker 2008), aiming to answer the 6 researches questions and accomplish the 2 objectives. The key aspects considered in selecting this research`s methods are listed at the table 2 (Amended from Wisker 2008) below. A result-oriented approach similar to Wisker (2008) has been taken which begins with defining the targeted results in order to narrow down the methodology selection.

| Objectives | Method | Information availability | Data Type | Sample Size | Statistical significance | Selected method |
|---|---|---|---|---|---|---|
| Objective I | Method I -A | Yes | Qualitative | Low | Low | **Literature review** |
| | Method I -B | Yes (Open source data) | Quantitative | High | High | **Trend analysis (Linear regression & Extrapolation)** |
| Objective II | Method II | Yes | Qualitative | Low | Low | **Literature review** |

**Table 2. Reserch Methodology Selection (Amended from Wisker 2008)**

**Method I -A**

As the previously conducted researches are still not aligned in the concluded impact of OTTs on MNOs. Method I-A investigates an extensive range of sources including researches that previously examined the impact of OTTs on MNOs in addition to whitepapers and telecom journals excerpts to provide in depth evaluation of the OTTs impact. This method aims to pull previous researches together to add synopsis and breakdown the researchers` conclusions. To achieve this objective, a qualitative method has been selected based on literature review as per Kothari (2004).

**Method I -B**

While method I-A concentrates on collecting, analysing and generating a non-numerical data, method I-B uses actual numerical data in order to support the literature review with a referenced financial open source data collected from organizations, whitepapers, telecommunications journals excerpts and recognized statistical telecom bodies including GSMA, Informa and Ovum, in addition to the MNOs` annually and quarterly open source released financial reports. As the research targets to study the global telecom industry, GSMA (2020b) reports have been used that includes a definitive source of mobile industry data which covers every MNOs all over the world. Quantitative methodology was adopted based on regression and extrapolation targeting to answer the research questions articulated above. Linear regression of phase I actual revenue values followed by extrapolation to forecast the baseline expected revenue of phase II and III, is followed in order to statistically evaluate the hypothesized revenue`s nonlinear trend over the years following Lavrakas (2008), Pickard (2007) and Grove (2004). Where linear regression is a process applied in phase I to derive a single line that best fits the given set of revenues data points over the years (X and Y). The X values are the year, and Y values are variable which is the revenue. Linear regression using least square values methodology is used to draw the revenue trend line that mostly fits phase I non-linear revenue trend. Most graphing calculators including Microsoft Power BI, QLIK and Excel can draw this line.

Once the revenue line is visualized in phase I, an equation will be derived that draws this revenue line. In order to predict a value of Y which is out of our known values set, Linear extrapolation method will be used at which a value X (Year) is given to get resulting value Y (Variable) to expand the trend line in phase II and III. Linear extrapolation is the process at which value of X being evaluated into the equation of the line of best fit to get a resulting Y value. In statistics, extrapolation is a process followed to estimate unknown values beyond a given variables in distinct range as defined by Seber and Lee (2003) and Montgomery et al. (2012). The accuracy of the forecasted data is conditional on the continuity of the same situation in the future, as this method tries to predict the future data based on historical data (Byjus 2018). Extrapolation can be applied by extending the known series or values beyond the know area to the unknow area. It is applied in this research by drawing a tangent line among the at the endpoint of phase I and that will be extended till Phase II and Phase III Seber and Lee (2003). In this study, the known area is phase I and the unknown area is phase II and phase III. Extrapolation concept can be applied on different disciplines including telecommunications with some categorical data. The expected accuracy from applying this method in this research is highly supported by Seber and Lee (2003) explanation that this method is suitable for any linear function and in this report phase I trend will be converted into linear trend using the regression. Also, Montgomery et al. (2012) explained that when the unknown values to be predicted is not too far from the given data, linear extrapolation provides an accurate result, which is the case in this research as Phase II and Phase III are not far from Phase I data period.

**Method II**

Several researchers have examined the OTTs impact on MNOs and accordingly recommended strategies to overcome this impact. Method II in this research will focus on analysing and gathering the different recommendations of researches and authors together to narrow down the strategies into 3 holistic main strategies which helps to align the different strategies. Then each strategy recommendation will be analysed and assigned one of the 3 holistic strategies which can fit the most into it. The idea is to quantify the recommendations orientations to get the most recommended strategy which will support the final recommendation of this research. Literature review methodology has been selected for this objective, backed by data analytics.

## III  Literature review

### III.I  OTTs are threat or opportunity for MNOs

Whereas the actuality is that the global revenue growth rate of MNOs is declining according to Ovum (2018) and GSMA (2020b), the researches` dispute is still ongoing on examining how OTTs contributed in this declining MNOs` trend. Various research papers have been conducted to isolate and determine the OTTs` effect on MNOs, nevertheless they have been concluded in two reverse directions:

First direction adopts that OTTs have formed a genuine threat on MNOs which reduced their revenues and jeopardized the MNOs` current business model. BMI (2018) listed OTTs in the threat`s list after conducting SWOT (Strengths, Weaknesses, Opportunities, And Threats) analysis on USA`s (United States of America) telecommunications market as OTTs have shifted MNOs` traditional SMS and voice revenues to their own applications. Similarly, Fowora et al. (2018) concluded in their study that OTTs have harmed the MNOs business as the low-cost digital content they offer to consumers has turned the consumers to prefer using OTTs services that utilizes the MNOs networks without any lease agreement or policies to regulate them. However, Fowora et al. (2018) underlined that MNOs registered substantial losses due to OTTs, they have not quantified this loss, so the significance of their study is low. Equally, Joshi et al. (2016), without a solid evidence, have directly pointed out to OTTs of being the major contributor of draining MNOs` revenues and ARPU. With non-empirical study, they propounded revenue sharing models between MNOs and OTTs. Also, Crawshaw (2017) unswervingly underlined OTT providers as a factual threat on MNOs, likewise Kwizera et al. (2018) and Barclay (2015) who acknowledged that the OTTs are the essential reason of disrupting the traditional business model of telecommunications so the regulatory have to take actions to resolve the issue. With more offensive approach, Sujata et al. (2015) proposed blocking the OTTs by MNOs in order to contain OTTs` threat, nevertheless in their study they failed to illustrate enough the OTTs` threats.

The dispute continuing, in the quantitative study done by Whitwell (2017) to scale the OTTs risk on MNOs, the survey results showed that 43% of western Europe considers OTTs as threat and central Asia region considers it as threat by 33%. However, Whitwell (2017) survey is stronger than Crawshaw (2017) survey as they included more participants from OTTs, their survey results were not biased enough towards certain direction, as it got nearly equal responses number for threat or opportunity debate. Generally, this researcher`s orientation censures the reason of the MNOs` revenue loss to the regulators of not controlling OTTs in the market with firm actions implementation. From their side, regulators, decided not to regulate OTT services and to be technologically neutral. So Cataldo (2016) in his research, accused regulators of harming the growth of mobile telecommunications industry and urged them to apply strict laws and regulations against OTTs. While most of researches focused on MNOs and regulators, Linnhoff-Popien et al. (2017) have examined OTTs position in this dispute and concluded that OTTs are substantially rejecting any specific laws and policies particularly in USA.

Contradictory, the second direction of researches identified the OTTs as an opportunity for MNOs. In their survey results, Schneider and Hildebrandt (2017) indicated that MNOs are likely to profit from the OTTs uprising trend as they are resulting into an increase of consumers' willingness-to-pay for MNOs mobile access. Strok et al. (2016) and Esselaar and Stork (2018) claimed that OTTs represent a considerable opportunity for MNOs which resulted in obvious upsurge of mobile users, smartphones and data traffic, which increased the MNOs` revenues from the induced data revenues generated by OTTs usage, that compensated the inherent decreases in the traditional voice calls and SMS revenues. Though, the revenue growth trends used for some countries in their research are not aligned if compared to GSMA (2020b) global MNOs market data. Furthermore, despite Stork et al. (2016) and (Esselaar and Stork 2018) showed an annual increase in revenue growth trend for MNOs to support the hypothesis that OTTs are opportunity, their reported revenue growth is less if compared to the period from 2009 till 2014. Esselaar and Stork (2018) blamed the declining revenues of the MNOs on the insufficient network coverage of the 3G and 4G mobile networks or on obsolete operating conditions or outdated voice and SMS only business models ; MNOs with wider mobile networks coverage that have invested in network upgrades or modernized their business models , have succeeded in monetizing the significant increase in traffic, hence compensates voice and SMS revenue losses.

Esselaar and Stork (2018) appealed the MNOs to envisage OTT services as a sufficient tool to wheel up data traffic growth and to enlarge their subscribers` base, and the regulators to stimulate MNOs` investment into 5G coverage instead of protecting MNOs from OTTs. Similarly, Trask (2018) supported this direction of considering OTTs as an unique opportunity to MNOs and emphasized that OTT providers are not yet subject to a firm regulation, which is aligned with Monarat and Hitoshi (2019) who concluded that applying regulations on OTTs like MNOs is not preferable solution. Likewise, Schneider and Hildebrandt (2017) raised a caution to policy makers and regulators not to employ similar rules on OTTs as MNOs as their functionalities are different and warned from consequences that might arise on consumers. In the same way, BEREC (2015) accentuated that OTTs are opportunity that enables MNOs to increase their data revenues and attract new customers as well as reduce churn, hence MNOs have enough incentives to seek partnering with OTTs. This orientation of researches generally calls to protect OTTs from strict regulations, and attacks the MNOs of not upgrading their strategies, therefore this direction promotes a cooperative strategy that MNOs must take forward with OTTs to increase their revenues.

### III.II Impact of OTT on MNOs

Wether it is threat or opportunuty, there are solid variations occurred on MNOs` main sources of revenus due to the emergence of OTTs. Detecon (2017) cited by Krämer and Jalajel (2019) reported that only Facebook, Instagram and Whatsapp have outgrown 8 major MNOs including Vodafone, Deutshe Telecom and Verizon, and in regards of the number of subscriptions; MNOs have got 2.11 billion subscribers compared to 2.45 OTT users). However, this comparison is not valid as the 8 MNOs` footprint is limited if compared to those 3 OTTs whose footprint is global. The increase in mobile user's number happened due to the extensive infrastructure offered from MNOs; MNOs overall have spent billions of dollars to make the infrastructure. Additionally, as countries` regulations, MNOs have to purchase the frequency band license (Heinrich, 2014). Knoben (2015) described OTTs as diverse universe; OTTs operations are not profitable but they target to monitze their investments by advertisements. Heuermann (2019) elaborated that OTT providers are using a "Freemium" model which is a recipe of "free" plus "premium", this business model became an overriding model for all new OTTs. At which, customers get the basic offered features at zero cost then can access advanced functionalities with a subscription fee. The OTT companies can afford using this model as they are not paying any frequency bandwidth nor regulatory nor licenses fees. OTTs just using MNOs` networks who are paying all of those (Heinrich, 2014).

As shown at table 3 (Amended from Knoben 2015,p.17 and Bhawan and Marg 2015, p.6 ) similar to Minges and Kelly (2018) but with adding the MNOs colomn in this reserch to compare the revenue sources. The OTT providers are categorized into 4 calsses based on the offered services. The first two calsses are considred as less impacting on the MNOs revenues, commerce OTTs such as Amazon and Paypal that offeres shopping and finance services and their revenue srouce is from transactions, and the social media such as Facebook and Twitter with revenues from advertisment and premium subscribtions. Other two OTT calsses are, OTT communications that offeres voice and messaging services and directly overlabing with MNOs` offered services and lastly OTT media which offers content services such as Netflex and Youtube. The last 2 OTT classes` services are directly overlabing with MNOs offered traditional services and hence the revenue sources of LDC,SMS, voice calls and data, which will be deeply explored below :

| OTT | | | MNO |
|---|---|---|---|
| OTT Classess | Example | Revenue source | MNO revenues source |
| OTT Commerce | Amazon, PayPaal, eBay | Transaction based | LDC,SMS,Voice calls ,Data |
| OTT Social Media | Facebook,Twitter,LinkedIn | Advertisement, subscription for premium services, free services | |
| OTT Communications | Skype,Whatsapp,Viber, Facebook messenger. | Advertisement, Subscription for premium servies, free services | |
| OTT Media | Youtube, Netflex,Spotify | Advertisement, subscription for premium services , free services transaction based. | |

Table 3 The OTT vs MNOs Business Models (Amended from Knoben 2015, p.17, and Bhawan and Marg 2015, p.6)

**(LDC) Long distance calls**

The OTTs` impact has been intense in what has been long lasting as a major traditional profit centre for MNOs which is the international voice calls. Farooq and Raju (2019) elucidated that the international voice calls` revenues for MNOs are extremely decreasing as a result of the quick emergence of OTTs` applications such as Messenger, Skype and WhatsApp. Only Skype handled a third of international voice calls traffic globally at 2012 as reported by Goldstein (2013). Joshi (2015) further explained that the offered services of OTTs with high flexibilities and low or zero cost in comparison to the MNOs` cost, is the main reason that users are preferring to use OTTs` international calls. The international conversational voice minutes number was increasing for MNOs till 2012 where it commenced to decrease afterwards, on the other hand the voice minutes carried by OTTs started to increase since the beginning and exceeded the MNOs minutes number at 2016 (Minges and Kelly 2018). As a result of this, Stroke et. Al (2016) mentioned that some MNOs has attempted to prevent customers using OTTs such as Skype as example to protect their international calls revenues.

**(SMS) Short message service**

The traditional MNOs` service of sending a message from the mobile device is based on CS (circuit switching) networks technology and it is called SMS. The MNOs` SMS numbers are decreasing over the years and being replaced by PS (packet switching) networking technology based on IP (Internet Protocol) (3GPP 2015), where all the OTTs are using to deliver their content over MNOs` networks freely. Those variations were followed by a sharp decline in MNOs` conventional messages services (SMS) numbers, which used to be sent exclusively over their networks. This decline is backed by the analysis done by Ofcom (2016) on 17 major telecommunications markets, where it was found that the number of SMS messages reduced in 14 out of 17 countries. Specifically, Spain had a decrease of 36% in SMS number if compared to 2015. Similarly, IPcarrier (2013) reported that SMS usage in China has declined nearly 11% in 2012 and OTT is the reason. Also, Heuermann (2019) reported that MNOs encountered 49.6 $ billion revenue loss from SMS at 2014. Saghaeian (2015) has emphasized the MNOs` SMS traffic decrease with the reported quarterly figures and concluded a year to year decline in SMS traffic by 7% starting 2013. This decline has started at 2011 after the global SMS traffic reached its peak of 7.4 trillion messages at this year meanwhile the OTT messaging has been increasing rapidly. Reactively, few MNOs started developing their own OTT applications and using IP messaging same as OTTs, but it is way far from the achieved OTT messaging numbers as per Krämer and Jalajel (2019) and ITU (2018).

The decline of the MNOs SMS trend was due to the huge shift from users towards OTT messaging, there are major reasons for this shift that is illustrated in the comparison held at table 4 (amended from Farooq and Raju 2019, p.7) to explore the major 13 features availability Of MNOs` and OTTs selected top 4 providers, similar idea to Farooq and Raju (2019) but with different OTTs selected ,adding extra features and adding a total score at the end based on overall features availability to conclude the overall difference in this research. As a total score, the traditional SMS having only 1 feature if compared to 12,11,11 and 11 features from Telegram, Whatsapp, Facebook messenger and Skype respectively. This huge difference of features availability assures the diversity and flexibility of the OTTs offered features in comparison to the limited MNO outdated features that don't fulfil the consumers requirements. This result shows that MNOs have to develop their offered services to cope with the customers' demands and to be able to compete with OTTs.

| Service Features | MNO services | | OTT services | | | | | | | |
|---|---|---|---|---|---|---|---|---|---|---|
| | SMS | | Telegram | | WhatsApp | | Facebook messenger | | Skype | |
| Text | ✓ | 1 | ✓ | 1 | ✓ | 1 | ✓ | 1 | ✓ | 1 |
| Audio call | × | 0 | ✓ | 1 | ✓ | 1 | ✓ | 1 | ✓ | 1 |
| Video call | × | 0 | ✓ | 1 | ✓ | 1 | ✓ | 1 | ✓ | 1 |
| User stories | × | 0 | ✓ | 1 | ✓ | 1 | ✓ | 1 | × | 0 |
| Open source | × | 0 | ✓ | 1 | × | 0 | × | 0 | × | 0 |
| Characters | 160 | 0 | Unlimited | 1 | Unlimited | 1 | Unlimited | 1 | Unlimited | 1 |
| Group chat | × | 0 | ✓ | 1 | ✓ | 1 | ✓ | 1 | ✓ | 1 |
| Emoticons | × | 0 | ✓ | 1 | ✓ | 1 | ✓ | 1 | ✓ | 1 |
| Photo sharing | × | 0 | ✓ | 1 | ✓ | 1 | ✓ | 1 | ✓ | 1 |
| Video sharing | × | 0 | ✓ | 1 | ✓ | 1 | ✓ | 1 | ✓ | 1 |
| Location sharing | × | 0 | ✓ | 1 | ✓ | 1 | ✓ | 1 | ✓ | 1 |
| Free app download | N/A | 0 | ✓ | 1 | ✓ | 1 | ✓ | 1 | ✓ | 1 |
| MSID number call | × | 0 | × | 0 | × | 0 | × | 0 | ✓ | 1 |
| Total score | 1 | | 12 | | 11 | | 11 | | 11 | |

**Table 4. MNO vs OTT messaging services (Amended from Farooq and Raju 2019, p.7)**

**Voice calls**

Traditional voice communication service was the major revenue source for MNOs (Czarnecki and Dietze 2017). But, with the emergence of smartphones and mobile broadband, mobile subscribers have been significantly shifting to using OTTs` voice applications such as Line, Viber, Whatsapp, etc instead of the MNOs traditional voice services. MNOs encountered 22.2 Billion $ in fixed telephony voice and 31.9 billion $ in mobile telephony due to the OTTs (Heuermann 2019). As per Kraemer (2017), only Whatsapp hits 100 million calls per day. Accordingly, the MNOs revenues from voice calls was steadily declining (Lotz and Korzunowicz 2019). Buzdar et al. (2016) and Farooq and Jabbar (2014) justified the consumers shift to OTT services as it is much advanced and widespread than what MNOs services offered. MNOs are offering affordable internet bundles including cheap voice calls, but customers are buying them to use OTT services. Buzdar et al. (2016) warned that if MNOs continue in offering such bundles that empowers the OTTs, they will completely lose their messaging and voice revenues. This has forced MNOs to invest in launching 4G and 5G new radio access technologies in order to maximize their revenues from alternative streams (Czarnecki and Dietze 2017). Unlikely, this investment has accelerated the erosion of their voice calls revenues as the rapid rollout of new technologies coverage led to increased service speed and decreased latency making the OTTs` voice quality comparable and even better than MNOs` offered voice service. This has strengthened the OTTs penetration into the market and empowered the OTTs voice applications which was an undesired consequence for MNOs (Lotz and Korzunowicz 2019) and (Czarnecki and Dietze 2017).

Consequently, the MNOs were on the brink of turning into a data pipes` supplier for OTT providers who are running their applications over their networks without incurring any cost. While the MNOs` infrastructure required to offer their traditional voice services (CS) circuit switching is different than the infrastructure required for OTTs` voice services VOIP (Voice Over Internet Protocol) which based on PS data network infrastructure 3GPP (2015). MNOs stuck in a situation that they have to maintain 2 different sets of expensive of infrastructures for their own CS voice and PS data for OTTs` voice applications (Lotz and Korzunowicz 2019), without any participation from OTT providers in this infrastructure investment (Czarnecki and Dietze 2017). To cope with this change, MNOs has introduced VOLTE (Voice Over LTE) which is a voice service carried over LTE data networks and it is quite similar to the OTTs voice services infrastructure requirements with few extra nodes such as IMS (IP Multimedia

Subsystem) (3GPP 2013). In this situation the MNOs have only single infrastructure for both their voice services and OTTs voice services. As the traditional CS voice services were continuing in declining, also due to the need of more spectrum to handle the enormous voice and data traffic, the MNOs have started decommissioning their legacy expensive 2G and 3G networks (Lotz and Korzunowicz 2019). Minges and Kelly (2018) reported that MNOs voice revenues in 2015 was 46% out of overall revenues, however it looks high, but it has sharply decreased as compared to 66% in 2010. Furthermore, ETTelecom (2020) anticipated that MNOs` voice revenues most likely to drop by 45% at 2024 compared to 2019. As in 2019 voice revenues recorded 381 $ Billion and will decrease to 208 $ Billion in 2024 and this is due to the increase of OTTs voice providers number by 88% in the next 5 years.

OTT providers have cannibalized MNOs` messaging and voice services after luring MNO`s customers with their free messaging, voice and also video functionalities. Such digital-savvy OTT companies - Beyond the "free" cost - are offering innovative VAS (value-added services) in order to complement their services to customers. The innovative VAS pays off; in 2011, fixed voice, mobile voice and messaging services that offered by OTTs, accounted for 11%,2% and 9% of revenues respectively. After 7 years only, in 2018, the numbers increased to approximately 50, 25 and 60 % (Meffert and Mohr 2017).To analyse the reasons that urged the customers to use OTTs voice applications instead of MNOs` voice, a gap analysis held at table 5 (personal collection) including the main 7 features based on consumers demands from MNOs and 4 selected OTTs.

| Category | Telecom product | | OTT services | | | | | | |
|---|---|---|---|---|---|---|---|---|---|
| Features | Voice | | Viber | | WhatsApp | | Facebook messenger | | Skype |
| International Call to any mobile number | ✓ | 1 | × | 0 | × | 0 | × | 0 | ✓ | 1 |
| Call to Landline number | ✓ | 1 | ✓ | 1 | × | 0 | × | 0 | ✓ | 1 |
| Group Calls | ✓ | 1 | ✓ | 1 | ✓ | 1 | ✓ | 1 | ✓ | 1 |
| Price | ✓ | 1 | ✓ | 1 | ✓ | 1 | ✓ | 1 | ✓ | 1 |
| Mobile Data required | Not required | 1 | Required | 0 | Required | 0 | Required | 0 | Required | 0 |
| Min Phone requirement | Feature | 1 | Smartphone | 0 | Smartphone | 0 | Smartphone | 0 | Smartphone | 0 |
| Call Reliability | ✓ | 1 | × | 0 | × | 0 | × | 0 | × | 0 |
| **Total score** | **7** | | **3** | | **2** | | **2** | | **4** | |

**Table 5. MNO vs OTT voice call services (personal collection)**

Unlike the SMS, the overall score of voice applications from MNOs is 7 compared to 3,2,2,4 for Viber, Whatsapp, Facebook messenger and Skype respectively. This explains the surviving trends for the voice calls over the MNOs that will be analysed at figure 10 p.26. On the other hand, the voice revenue is not similarly increasing with the increasing voice calls, the reason behind this is that is the offered free voice calls minutes by MNOs to the customers on top of their purchased data packages to use the OTTs as a prime demand.

**Data Traffic**

The rapid uptake of OTT applications in global telecommunications market has resulted in exponential increase in data traffic. According to research conducted by Meffert and Mohr (2017) in January, on average, this generation spends about 315 minutes online every day. Verma (2014) highlighted that MNOs are facing a tremendous change in their stagnating market due to the usage behavior change with the smartphone penetration, as it generates 14 times data volume more than if compared to the legacy feature phones (Czarnecki and Dietze 2017). Mobile users' numbers are significantly increasing, GSMA (2020a) reported 5.2 billion mobile subscribers compared to 6.3 billion reported by EIU (2020), but EIU reported non-unique subscribers' numbers. One can only imagine the volume of digitized data streaming over MNOs networks across the world those users are generating! This increasing trend causes severe congestion in mobile networks due to the surge demand of video content. Minges and Kelly (2018) identified MNOs` challenge as, voice and text, traditional revenue earner for MNOs are declining as users shift to using OTT services which the data traffic volumes to pop up; in 2011 the smartphones` data traffic surpassed the voice traffic. Based on Cisco (2018) Visual Networking Studies, the mobile data volume recorded 5 Exabytes per month at 2013 and has grown by 61% reaching 15.9 Exabytes at 2018, specifically, video traffic increased from 0.6 Exabytes to 9.1 Exabytes. This addresses the key challenge for MNOs with their limited capacity backhaul network represented in the intermediate links between radio access and core networks, which is close to saturation as explained by Nexans (2020).

The growth of data traffic has pushed MNOs to invest more in network capacity enhancement by requiring more spectrum, but also deploying more infrastructure in small cells. This trend of increasing data usage and the serious threat of OTT services will force MNOs to adopt a new business model. Also, MNOs would have to consider the growth in data traffic as another business opportunity due to proliferation of OTT communication services. While MNOs witnessed a loss of voice and SMS revenues, they have noticed a dramatic increase of data revenue. Peterson (2015) warned that without the transformation of their business model, MNOs` would be considered as dump pipes instead of traditional broadband MNOs. Particularly, video content OTTs provide different challenge, these include providers of television and films such as Netflix throughout subscription services and Youtube with free services, in which the video demand is increasing by customers (Peterson 2015). Similarly, (Sandvine 2015) highlighted that OTTs are not only competing with MNOs that providing video content services, but also, they are responsible for an increased substantial traffic volume that goes over MNOs` networks. As the example given by Sandvine (2015), Facebook properties` (Instagram & Whatsapp) and Googles` (Youtube, Google cloud and Google market) are accountable for 60% of total MNOs` traffic in whole Latin Americ. And globally, top 10 OTT providers consumes 71% of MNOs` global traffic. However Krämer and Jalajel (2019) claimed that OTT services increases the MNOs` core revenues as it increases the MNOs` data usage, voice and messaging ,hence it enlarges the MNOs` footprint, the claim was not right, as the OTTs services only increase the data usage not voice and messaging as the voice and messaging used by OTTs are based on PS data service which increases the data only as per 3GPP (2015). The reality even, that it decreases the voice and messaging of the MNOs as it competes directly with them by their services.

### III.III      MNOs strategies to overcome OTT impact

When the OTTs appeared on the telecommunication market, MNOs` were unsure how to react while being threatened. MNOs have pursued several responses and strategies to deal with the challenges arises from OTT providers, ranging from blocking to partnering (Czarnecki and Dietze 2017) and (Knoben 2015), or ranging from denial to emulation strategies as stated by Krämer and Jalajel (2019, p.211). Blocking OTT voice services is one of the strategies that MNOs have used. On the other hand, some MNOs chosen the partnership with OTTs instead of blocking them, furthermore some MNOs have developed their own OTT-Likewise services and added to their digital business divisions. Those 2 approaches represent minority of cases so far. Particularly, developing own OTT service is a strategy that requires higher maturity level in digital business areas and still in its early stages. Meanwhile, the ongoing OTTs` market developments are increasing the pressure on MNOs, giving them a very narrow opportunity window to conceive an appropriate and effective response strategy (Czarnecki and Dietze 2017). Similarly, Minges and Kelly (2018) elaborated that MNOs have developed several responses to OTTs. They have argued to regulate OTT providers which offers voice and messaging services same way MNOs` are regulated.

The outcomes of those strategies were assorted, at best was to slow down the development of the OTT offered services, however, not to terminate it (Krämer and Jalajel 2019). Crawshaw (2017) has conducted a quantitative survey on MNOs to examine the MNOs` strategies followed towards the threat from OTTs. Despite Crawshaw`s (2017) survey high sample size of 119, the OTT providers participants were only 8%, more participants from OTT providers should have been involved to get them into the scene and establish a closed loop solution. The popular response in this survey was 42% as partnering with OTTs to offer a combined bundled service. On the other hand, a brave 24% chosen a competition strategy against OTTs by developing their own OTT applications. And the rest of participants chosen a neutral strategy. Unsimilar to Crawshaw`s (2017) survey results, Czarnecki and Dietze`s (2017) claimed that MNOs mostly implemented defensive strategies.

In order to resolve this dispute and measure the significance of different proposed strategies, several researches proposals of the MNOs` strategic responses to OTTs have been aggregated and examined in order to conclude the most appropriate strategy in this research. The analysis shows that all researches' proposals can be categorized into 3 major holistic strategies (Compete, Partner and Accept) described as following:

## *"Compete" Strategy*

### 1. Seeking for regulatory support

The spontaneous reaction from MNOs was tackling the OTTs` threat by applying to authorities to change regulations. Requesting permissions to ask OTT for the cost of the traffic handled by their networks. Telefonica, KP and Italia telecom are examples of MNOs who have applied to European commission for regulating the OTTs (Gómez-Uranga et al. 2016). MNOs are calling to charge OTTs for using their networks, seeking to monetize OTT network traffic. Applying "sysball" principle as defined by Knoben (2015), which is the paradigm shift from 'content is king' to 'access to end customer is king' as the MNOs only providing access not content as OTTs. Apparently, there is a market and market entry conditions` imbalance between the OTT providers and the licensed MNOs as there is no proper OTTs` regulation yet. Regulatory requirements and limitations determine the MNOs` business models, on contrarily, OTT providers are free of such limitations. The market setups currently have not yet adapted to this new competitive situation (Knoben 2015). Table 4 (Amended from Knoben 2015 and Czarnecki and Dietze 2017 P.37) explores the comparison between licensed MNOs and OTT providers, 7 restricted regulations are applied on MNOs while OTT providers are free from those restrictions.

| Regulation | Licensed MNOs | OTT Providers |
|---|---|---|
| Licensing | License required for market entry and frequency band usage | No requirements for license |
| QoS (Quality of Service) | Regulatory set targeted SLA (Service level agreement) on the MNOs in the license, which is subject to fines, if not achieved. | No QoS requirements or commitments |
| Interconnectivity | Interconnectivity is mandated in the license | No interconnectivity requirements |
| Universal services | Must follow global services obligations | Not following universal service regime |
| Customers` protection | Subject to enforceable consumer protection policy | No enforcement power |
| Legal interception | License condition | Depends on country |
| Taxes | Must obey country`s taxes regime | Depends on service type |

Table 6. MNOs vs OTT Regulations (Amended from Knoben 2015 and Czarnecki and Dietze 2017, p.37)

From a technical point of view, OTTs provide consumers with some ease of communication and quality of service, to some extent. However, the fact remains that there is absence of established rules of the game, especially, issues related the competition with the traditional MNOs, supervision of OTT activities, personal data usage conditions, data traffic monetization, taxation issues, provision and operation obligations. The regulatory framework absence is linked to the lack of a solid OTTs definition, which has a direct impact on obligations governing telecom activities as well as business relationships (BEREC 2015). The regulators position is not supportive to MNOs; accordingly, MNOs have to think further to deal with the OTT threats.

### 2. Blocking the OTT Services

Several MNOs have determined to attack the OTT services by prohibiting their subscribers from using OTT applications over their networks, targeting to rebuild OTT portfolio and integrate OTT service in product bundle. The blocking is achieved by merging technical and economic features that halt certain OTTs from using MNOs` IP services (Knoben 2015). The objective of blocking certain OTT service, is to secure the core MNOs` revenues by making the OTT service unavailable or unattractive. Du, Zain, Etisalat and STC are vital examples of MNOs who blocked OTT services as reported by Knoben (2015). Also, Spain and France previously have blocked OTT services when it offered overlapping services such as messaging and voice calla that impacted MNOs` revenues. Farooq and Raju (2019) justified blocking OTTs` services, as the OTT services were acting like MNOs but without paying regulation fees, neither fulfilling the responsibilities. Sujata et al. (2015) supported the OTTs` blocking strategy willing to defeat the OTTs threat on MNOs. But Sujata et al. (2015) have not addressed the negative consequences that might result from this action which was explored by Farooq and Raju (2019). Farooq and Raju (2019) suggested that MNOs can force OTTs to give a market share by blocking them for a short period, limiting the suggestion to the countries where there is no technology neutrality law which prevent such action. This law exists in USA and Europe as example where they can't ban services providers or even internet content. In countries where this law doesn't apply, MNOs can try the short period OTT blocking tactic, but on the other hand, the fear of losing is high from this tactic (Farooq and Raju 2019). Farooq and Raju (2019) addressed the negative impacts of blocking OTTs; as it decreases the revenues of MNOs, as customers mainly purchasing data bundles from MNOs in order only to use OTT services.

Farooq and Raju (2019) and Sujata et al. (2015) have given a very risky recommendation to MNOs to block the OTT service providers for a short period and ask OTTs for market share. While blocking could be considered at early 2010s, it is nearly impossible now to implement due to the pressure that will be created from the mobile subscribers as OTTs became a vital component of their daily life. All the researches that suggested blocking OTTs are irresponsible, the OTTs are vital tool in our life nowadays. As example, with the outbreak of pandemic COVID-19 corona virus at 2020, WHO (World health organization) is using WhatsApp as a communication channel with all the people in the world through health alerts that brings COVID-19 facts and required precautions to billions of people (WHO 2020).

### 3. Absorb the OTTs

Another strategy followed by MNOs was making OTTs ineffective from the customer perspective by absorbing OTT services. Saving money is the objective of consumers to choose using OTT services. In response, the MNOs offers large messaging and voice free bundles on top of the purchased data packages by consumers, aiming to get the consumers to use their traditional voice and messaging services back and abandon the OTTs applications. As OTT services usage is expanding and social media is getting very popular, Farooq and Raju (2019) urged that, offering such cheap bundles can help MNOs to get more habitual internet customers which leads to higher usage and revenues. Similarly, Stork et. Al (2016) supported this approach of unlimited voice call and text messages as MNOs can receive the desired ARPU when they set the price of the top-up packages.

### 4. Create own OTT or acquire them

MNOs are still exploring how they can create their own value-added OTT services and hop on the OTT trend. Krussel (2019) proposed an offensive or defensive strategy for MNOs by cannibalizing OTT services such as WhatsApp. MNOs to offer similar services to the OTT provided services is a possible strategy explored by Czarnecki and Dietze (2017) as well. However Czarnecki and Dietze (2017) considered launching MNOs` proprietary OTT service is the least developed option so far, HBO is a vital example opposing their claim; HBO Go is a vital example of how MNOs can become a branded strong destination for content as per Kastrenakes (2018) and Infopulse 2019). The owner of HBO, AT&T, are not stopping developments just here. They recently announced launching of several value-added services and products, including 5 broadband delivered OTT products. Similarly, UK's Three has launched content packages, offering OTT video and music subscription also a mix of TV channels. Another example is T-Mobile US that have launched a statewide trend offering zero-rated video services delivered over cellular network using Binge On application (T-mobile 2020). Equally, German giant Deutsche Telecom have offered its own IPTV OTT application Magenta, which is a virtual pay TV available on standalone basis (Omdia 2018).

On the other hand, some MNOs tried to compete with the OTTs by creating equivalent voice and instant messaging services. As example, Telefonica in 2012 has launched an instant messaging service targeting to repel Line and WhatsApp, nevertheless the app managed to get only 1 million downloads compared to 200 million by Line and 300 million logged by Whatsapp (Reuters 2014). Many MNOs have created their own OTT application, however most of them are offering video content services. In the previous decade, there have been few attempts by MNOs to deploy instant messaging and voice OTT apps. The MNOs focus should have been more on the voice and messaging OTT apps as it is what directly overlap with their major revenue streams.

The 4 explored approaches in "compete" strategy didn't bring the required outcomes for the MNOs, apart from the 4th approach of developing own OTT, but even this approach was not successful in the case of voice and messaging application. Generally, competing with OTTs which have taken full advantage of the major network externalities already, is a herculean task (Gómez-Uranga et al. 2016).

### "Partner" Strategy

As Jim Henson said "If you can't beat them. Join them". Some MNOs attempted the way of partnering rather than competing with OTTs aiming to expand their value and to benefit from their widespread. Currently there is turnaround in strategies that can summed up in cooperating with OTTs (Krämer and Jalajel 2019, p.211). MNOs who decided to partner with OTTs might take advantage from both OTT services and the OTT brand (Czarnecki and Dietze 2017). The price will remain the major drive of voice market so MNOs are using pricing levers in order to assure that their own voice services are appropriate to their consumers (Czarnecki and Dietze 2017) . MNOs targeted three major partnership`s objectives; to complement MNOs` own portfolio, OTT containment and securing high values segments by service differentiation. Krussel (2019) as well proposed the cooperation strategy however it was conditional; that only applies when there is an immediate little overlapping between OTTs` services and the MNOs services offered. Krussel (2019) highlighted the advantages of partnership as OTTs` service has got more of an additive character and wider spread, so it widens the MNO`s access to new customers` segments where the MNOs have a limited footprint, such as the entertainment services, Deezer and Spotify (Krussel 2019,p214).

Krussel (2019) urged MNOs to consider Netflix and Spotify as a big volume partners with the target to open new market segments and limit customer churn, however in the study, the positive impact of limiting the churn was not compared to the negative impact that might happen on the capacity due to traffic increase. Infopulse (2019) supported the cooperative strategies claiming that OTTs have got enough revenue and space to contain both MNOs and OTTs.

These days the trend of partnering with OTTs become ubiquitous among MNOs where they offer monthly internet bundles with specific OTTs (Farooq and Raju 2019). To be more promising than opposing it, MNOs have started to perceive the value of connecting with the efficiency and creativity of the OTTs. From 2012 till 2018, there were 943 partnerships have been established globally as shown at figure 2 (Amended from Krämer and Jalajel 2019, p.212). Following the advancements of the content providers such as HBO GO, Spotify and Netflix, the numbers have significantly grown; only in 2016, 61% of these partnerships were concluded with OTTs (Krämer and Jalajel 2019).  The MNOs benefit from the partnerships with OTTs by adding the OTT services to their own portfolio in order to strengthen their customer loyalty and 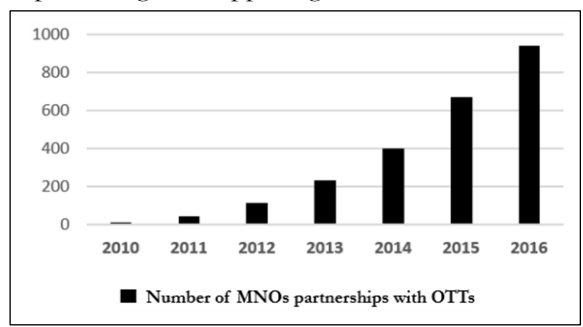

**Figure 2. Accumulative development of OTT partnerships (Amended from Krämer and Jalajel 2019, p.212)**

increase the MNOs usage, hence increases their footprint while stabilizing their revenues Krämer and Jalajel (2019). Additionally, new revenue streams can be created by innovative technologies offered and developed by the OTTs to the MNOs, so MNOs don't need to invest in developing their own OTTs (Krämer and Jalajel 2019).

For example, the Malaysian Maxis offers Whatsapp and Facebook free usage along with their daily, weekly or monthly bundles (Maxis 2020). Airtel is a vital example as well for partnering with OTT, when they partnered with Amazon Prime and offered it to their customers in a bundle contains free calling and unlimited data (Airtel 2020). Furthermore, Vodafone signed with Netflix and offered 12-month standard Netflex subscriptions to their customers (Vodafone 2020). Mobaily, Viva, Ooredoo, STC and Nawras are also examples for parenting with OTT providers (Knoben 2015). With the increasing partnerships and offered bundles to users, Infopulse (2019) explored the consumers perspective of partnerships through a survey that showed 51% of consumers wished to access all the video content they watch with a monthly fee regardless where it is published.

From the OTTs side, OTTs can benefit from the partnerships with MNOs to access their clientele and build a global user base. Particularly, they can start to learn about mass communications, customer touchpoints, customer acquisition from MNOs, which will benefit OTTs to gain market insights Krämer and Jalajel (2019). On the other hand, the partnerships with OTTs entail risks to MNOs as well. It accelerates the market entry of new potential competitors in telecommunication market which may result into further shrinking to MNOs market shares and losing customer relations. For the OTTs the risk resides in the connectivity dependency they have on the MNO`s infrastructure. The existence of MNOs from the cooperative venture might shut down OTTs operations immediately and significantly harm OTTs` revenues (Krämer and Jalajel 2019). From one perspective MNOs are confident that their services may gain popularity among consumers, however from the other side, MNOs are not confident as their own services may get marginalized by OTTs, which push the MNOs to continue seeking other strategies.

## "Accept" Strategy

If not competing neither partnering, some MNOs have preferred a hands-off strategy to any application that may upsurge in the data traffic usage, including the OTTs` services as well. Such MNOs are convinced that that communications` services non-occasional nature such as messaging or voice over IP leads to intense incentives for consumers to purchasing upgraded data plans. Those MNOs are trying to monetize the data traffic generated in their network by OTTs (Czarnecki and Dietze 2017). Jalajel (2018) described this approach as the Zero strategy of doing nothing towards OTTs, accepting the nature of OTTs proliferation, OTTs are increasing rapidly, and MNOs must accept this fact. But in this research "Accept" strategy developed further besides the "zero-approach" to include that MNOs have to develop alternative strategies to improve their business models, those strategies are not overlapping with the OTT services. Accept can be the most suitable strategy to cope with the OTT providers. As the OTTs proliferation is not just stopping here and more OTTs are developing every day. As evident in the online video by John Legere (2018) CEO of T-Mobile USA in "First in" CNBC interview, explained that 4G technology has created Uber, Airbnb, Snapchat and accelerated Amazon ,Facebook and Google, so with 5G innovation a whole new group of entrepreneurs are looking for creating enormous new OTT applications. Legere (2018) claim is aligned with Farooq and Raju (2019) conclusion that 5G is expected to evolve more in the world at 2020 with increase in internet speed from 1000 Mbps to 5000 Mbps, which will place an intensive pressure on MNOs to expand their infrastructure more, but again the main beneficiary of the new infrastructure will still be the OTT providers. 5G will boost OTT providers and Internet of Things (IoT) companies to operate better with this high internet speed, which will deteriorate the situation more for MNOs.

This means, applying the Compete or Partner strategies will be no longer effective in the near future as a new OTTs are coming to the market. Additionally, with the first 2 approaches implementations, MNOs revenues are still declining unlike the OTTs, as will be explored at data analysis section, so the strategies were not powerful enough to overcome the impact of OTTs. Accordingly, MNOs must Accept OTTs, but with developing parallel strategies to improve their business models. With the deep penetration of OTTs in people`s daily life and the difficulty to stop it, MNOs have left with no choice to accommodate the OTTs traffic over their networks and without OTTs paying for them. In this case MNOs should focus in optimizing their operational and capital costs, below are two approaches for CAPEX and OPEX saving.

### Network Virtualization to reduce CAPEX

Deploying the virtual RAN (Radio Access Network) reduces the network capital cost by 30%. Instead of the expensive hardware equipment's that MNOs need to build and expand their infrastructure networks to accommodate the OTTs traffic. MNOs must think in emerging the software-based networks or so-called virtualized networks which is 30% cheaper in the CAPEX requirements. In order to keep up with the constantly changing market conditions, a fully virtualized networks allows the shift away from the tightly coupled hardware and software traditional models to enable NVF (Network Functions Virtualization) technology. Krämer and Jalajel (2019) supported this approach emphasizing that more service providers are emerging to offer virtualized infrastructure, service orchestration and network function controllers. The Japanese giant, Rakuten, is a role model for the MNOs to optimize their infrastructure cost, Amin (2019) CTO of Rakuten mobile has taken a revolutionary approach to build world`s first end to end fully virtualized cloud native mobile network which commercially launched at April 2020. Amin (2019) emphasized that the journy they are embarking on in Japan enables the complete telecom infrastructure buildout transformation. In only 12 months, Rakuten has built this network without a significant CAPEX investment (Rakutentoday 2019). With a higher cost reduction achievement, Vodafone Group recently hit a 50% capital cost reduction of its core network functions by wrapping up the deployment of VMware's network virtual infrastructure at April 2020 across all its 21 business markets in Europe as reported by Robuck (2020).

**Automation to reduce OPEX**

Along with CAPEX reduction, Reith (2019) suggested that MNOs can focus on automation as operations automations helps MNOs to reduce their OPEX. Applying a heavily automated managed services with a zero human intervention will help to deliver better agility and QoS and improve customer`s experience in addition to the cost reduction. Krämer and Jalajel (2019) elaborated that automation processes scope that can include configuration, maintenance and troubleshooting and resource optimization. Automation of business processes was recommended by Hong and Dietze (2019) as well, as it brings multiple opportunities to MNOs in efficiency and cost reduction. Additionally, Schmitz et. Al (2019) concluded that robot process automation, stand-alone automation, Machine learning, and Artificial Intelligence as optimum approaches for MNOs` automation. Czarnecki and Dietze (2017) and O-RAN Alliance (2020) included automation and self-driving networks through learning-based technologies and new software systems in their transformational recommendations to MNOs stressing in the cost advantages that automation can bring to the MNOs` OPEX in addition to the ability to manage upcoming complex networks.

### III.IV Combined researches proposals

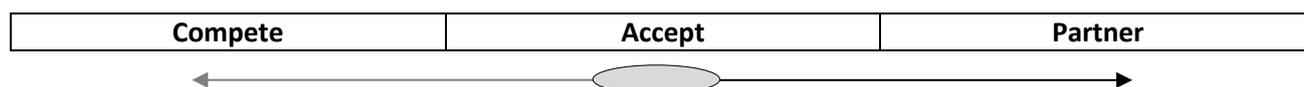

| Compete | Accept | Partner |
|---|---|---|

As explored in the previous sections, several researches have recommended different strategies for MNOs to react towards OTTs threat. Farooq and Raju (2019) listed 18 researchers' proposals in a table, however it was without any analysis or conclusion. More researches have been gathered here in this research up to 26, in addition to analytical approach to quantify the significance for each recommended strategy after analysing and mapping the recommended strategies into one of the 3 derived strategies in this research (Compete or Partner or Accept). In order to be able to give an overall bifurcation for all the recommendations and measure the recommendations significance. The 26 researches conclusions have been analysed and collected as shown at table 7 (amended from Farooq and Raju 2019, p.9). Figure 3 (personal collection) shows the bifurcation of the strategies with the percentage of the recommendations by the researches. 19% of the researchers recommended competitive approach, while 27% suggested cooperation and the popular recommendation were 54% supported the zero approach or the Accept strategy. This analysis shows the popularity of the accept strategy that MNOs have to take, in order to control the impact of OTT providers. By breaking down the 19% competitive approach results, results show 20% recommended blocking, 20% suggested regulatory support and rest 60% proposed alternative innovative strategies to indirectly overcome the OTT impact.

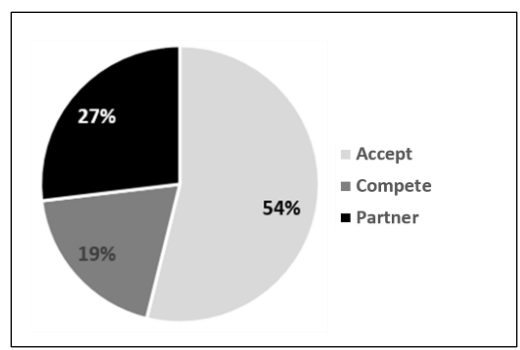

Figure 3. Research's recommendation breakdown - MNO strategies towards OTTs (personal collection)

| Author | Proposed strategy for MNOs to deal with OTT impact | Strategy |
|---|---|---|
| (Monarat and Hitoshi 2019) | According to the results, it is not preferable to regulate the OTTs similar to traditional telecom services | Accept |
| (Schneider and Hildebrandt 2017) | MNOs are likely to profit from the OTTs uprising trend as they are resulting into an increase of consumers' willingness-to-pay for MNOs mobile access. In fact, consumers use the two types of communications services very differently. | Accept |
| (BEREC 2015) | MNOs can attract new customers base and reduce churn by seek partnerships with OTTs. That will increase MNOs` data revenue and add value to their services. | Partner |
| (Farooq and Raju 2019) | Farooq and Raju (2019) has proposed an innovative idea to decrease the gap existing in the OTT services market. The MNOs can provide a combined platform that merge all OTTs enabling the interconnectivity between different OTTs. | Partner |
| (Infopulse 2019) | Now it's time for MNOs to invest more in building superior digital touchpoints with customers to improve customer experience | Accept |
| (Barclay 2015) | MNOs to push for regulatory option | Compete |
| (Kwizera et al. 2018) | MNOs to push for regulatory option | Compete |
| (Joshi et al. 2016) | MNOs to push regulators for revenue`s sharing models between OTTs and MNOs | Compete |
| (Krämer, 2019) | MNOs shall partner with OTTs which can resolve the ongoing problem | Partner |
| (Nooren et al., 2012) | MNOs get benefits from OTT players as it creates opportunities for telecom companies | Accept |
| (Borowski and Khurana 2019) | MNOs should focus on investing in 5G and monetizing its data traffic. | Accept |
| (Schmitz et al. 2019) | Robot process automation, stand-alone automation, Machine learning, and Artificial Intelligence are optimum approaches for MNOs` automation to reduce OPEX. | Accept |
| (Hong and Dietze, 2019) | Automation will bring multiple opportunities to MNOs in cost optimization, by business process automation, AI, M2M | Accept |
| (Wisselink and Schneider 2019) | MNOs to invest in IOT related products. | Accept |
| (Grineisen and Rehme, 2019) | MNOs to create their own OTT services applications. | Compete |
| (Glohr, 2019) | MNOs to focus on monetizing the data usage by their business consumers. | Accept |
| (Obernolte et al. 2019) | MNOs will gain revenues for the high OTTs penetration. | Accept |
| (Eberhard and Heuermann, 2019) | MNOs to focus in open access internet broadband deployment via the mobile wholesale network. | Accept |
| (Steingröver et al. 2019) | MNOs to urge authorities to regulate OTTs or MNOs to block the OTT services | Compete |
| (Heilen, 2019) | MNOs to cooperate with OTTs | Partner |
| (Weber et al. 2019) | MNOs to focus in cost optimization through managing the energy sources such as low power wide area networks as a game changer for the IoT. | Accept |
| (Lukowski et al. 2019) | MNOs to focus in machines connections that will reach 21 billion connections in2021 | Partner |
| (Heinemann et al.2019) | MNOs can compete focusing in digitization of Health, transportation, education | Accept |
| (Reith 2019) | MNOs can compete with a focus on Automation as cost optimization | Accept |
| (Korzunowicz 2019) | MNOs to help OTT companies in going through technology stability & change | Partner |
| (Krüssel and Göbel 2019) | Network expansion and optimization to provide and improve quality of OTT Services | Partner |

Table 7. Combined researchers' proposals for MNOs strategies towards OTTs (amended from Farooq and Raju 2019, p.9)

## IV  Data and analysis

### IV.I Major telecom trends analysis in the era of OTTs

Following the literature review, this section huddles in and build out the major mobile telecommunications industry trends that been impacted by the emergence of OTTs in order to examine the variations to support resolving the controversy about the OTTs impact on MNOs. Starting from 2020`s mobile telecommunications industry dashboard shown at Figure 4 (personal collection) that includes the key telecommunications` industry measures at 2020. The dashboard proofs that that telecommunications industry is a successful industry in the new millennium. Mobile users' numbers reached 5.2 Billion out of 7.77 Billion total population in the world with a penetration rate of 67%. The total revenue recorded in 2019 is 1.032 $ Trillion and ARPU 13.56 $. MNOs are delighted with this revenue and investing more in modernizing their networks to serve more users.

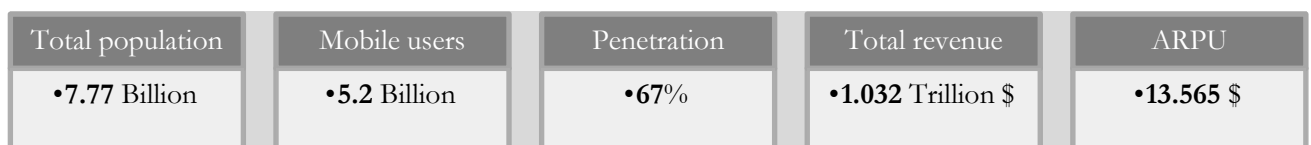

**Figure 4. 2020`s mobile telecommunications industry dashboard (personal collection).**

Although Figure 4. reflects the telecom industry success, the OTTs have significantly impacted those values. The severe impact of OTTs on MNOs represented in smartphones, LDC, SMS, voice calls and traffic, are analysed below:

**Smartphones**

Mobile types` connections shown at Figure 5(amended from GSMA 2020b) reproduces the first impact brought by OTTs to telecommunications industry. As shown at the figure, the feature phones which grant the customers to perform voice calls and send SMS only, were degrading significantly after reaching its peak at 2011. On the other hand, the smartphones which grants the customers to use OTTs, were rapidly increasing starting from 2008 along with increase in mobile-data only connections. Based on the capabilities of the feature phones in comparison to the smartphones, this trend indirectly emphasis a degradation in traditional voice and SMS confronted by a massive increase in OTTs usage.

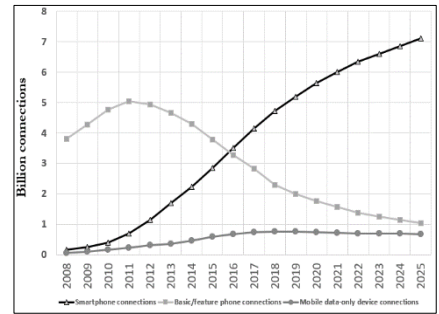

Figure 5. Mobile types` connections (amended from GSMA 2020b).

**LDC**

International conversational voice minutes traffic shown at Figure 6 (amended from Telegeography 2019) illuminates the impact of OTTs on MNOs` LDC number globally. The international conversational voice minutes number was increasing for MNOs till 2012 where it commenced to decrease afterwards, on the other hand the international voice minutes carried over OTTs started to increase since the beginning and exceeded the MNOs minutes number at 2016. By comparing 2017 vs 2012, the MNOs` international calls minutes number decreased by 17% from 617 to 514 million minutes. At the same period, the OTTs` international minutes number increased by 354% from 160 to 727 million minutes. The difference is remarkable and accentuates the negative impact of OTTs on MNOs. This highlights that the OTTs are aggressively replacing MNOs` LDC service which is a main source of revenue for MNOs along the years.

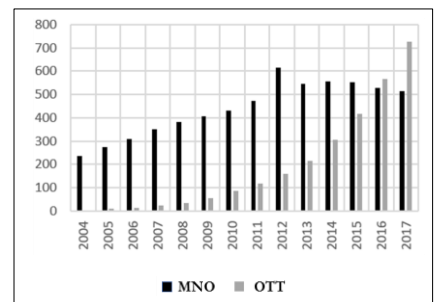

Figure 6. International conversational voice minutes Traffic (Million Minutes) (amended from Telegeography 2019).

**SMS**

Similar to LDC trend, the global SMS number trend is sharply decreasing as well, as shown at Figure 7 (amended from ITU 2018, p.7). The decrease of MNOs` SMS number is accompanied with exponential increase of OTTs` IP messaging number and a slight increase of MNOs` IP messaging. Unlike the LDC that started decreasing at 2012, the SMS trend of MNOs started decreeing 1 year earlier. By applying the same comparison of 2012 vs 2017 of LDC, the MNOs SMS number decreased by 17%, surprisingly it is same degradation value as LDC. On the other hand, the OTTs messaging number increased by 967% from 3 to 32 trillion messages, which is around 3 times more than the increase of OTTs` LDC. The difference in the OTTs` growth in SMS vs LDC support the analysis done at the literature review section, table 4, p.15 , that the OTTs are much advanced and closer to customers in the messaging service that's why their growth rate is 3 times higher than LDC.

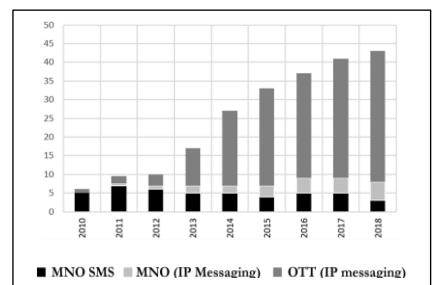

Figure 7. Mobile Messaging by service type (Trillion Messages) (amended from ITU 2018, p.7).

However, in the LDC there are limitations in OTT services that their applications are not capable of calling international mobile number if compared to MNOs, which limits their growth rate. The massive growth and advancements of OTTs messaging service ensures that the SMS is more danger to MNOs to be replaced by OTTs` messaging. By applying linear regression followed by extrapolation on the quarterly SMS number reported by GSMA (2020b) from all global MNOs as shown at figure 8 (personal collection). SMS will approximately vanish by Q4 2022. The SMS data reported in GSMA (2020b) is quarterly if compared to annual data reported by ITU (2018). The degrading trend of SMS number has resulted into a degradation in ARPU of messaging services globally for MNOs as shown at Figure 9. (personal collection) with a sharp drop at 2012 which is the turnover year when SMS number started to decrease after the increasing trend.

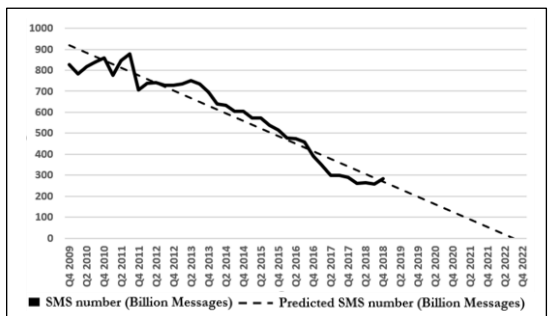

Figure 8. Global SMS number (Billion Messages) (personal collection)

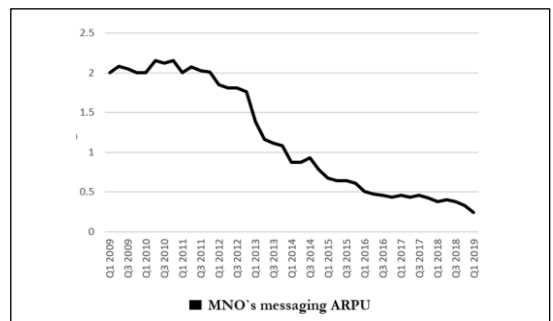

Figure 9. MNO`s messaging ARPU (personal collection)

**Voice calls**

While the LDC and SMS shows a sharp drop in usage by MNOs` customers, oppositely the voice calls trend is showing sustainability in increasing minute of use as shown at Figure 10 (amended from GSMA 2020b). The increasing trend of MNOs voice calls supports the analysis done at table 5, p. 17, which concluded that MNOs` voice calls are fulfilling the customers` need if compared to OTTs` limited capabilities of voice calls. One of the reasons is that OTTs voice calls have limitations to call fixed line numbers and MSIDN numbers. The trend shows that OTTs have not succeeded yet to decrease the MNOs` voice call services unlike the SMS and LDC.

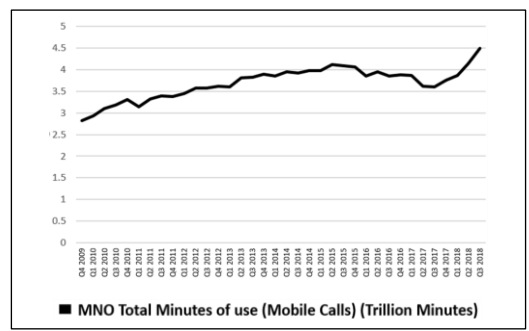

Figure 10. Total Minutes of use (Mobile) (Trillion Minutes) (amended from GSMA 2020b)

**Data traffic**

Figure 11. (amended from Ericsson 2019 and Cisco VNI 2017) shows the sustainable increase in global data traffic in download and upload services every quarter. The tremendous traffic increase is aligned with the increase of smartphones trend as in Figure 5, p. 24. Cisco VNI (2017) reported this increasing data trend and additionally predicted a 1100% growth in data traffic over smartphones by 2021 if compared to 2014. A remarkable increase in data traffic started at 2014, the reason behind this increase as justified by GSMA intelligence (2020a) is that the demand for data capacity grew quickly during the 4G era as consumers were using mobile broadband more. Accordingly, the main source of MNOs`

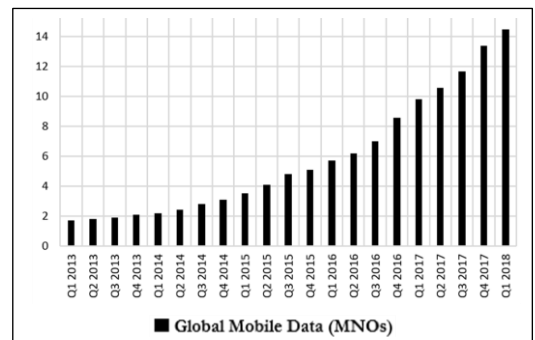

Figure 11. Data Download+Upload in Exabytes (Billions of Gigabytes) (Amended from Ericsson Mobility Report,2019,p.12, and Cisco VNI 2017)

revenues shifted from LDC, voice minutes and SMS to data traffic. In order to manage the increasing data`s demand, MNOs have thought to increase network capacity while reducing the network`s costs, which has been achieved through investments in new generations of mobile technologies, each of which is more efficient at handling data traffic.

The 5 major telecommunications` trends analysis done confirms that OTTs have significantly impacted and reshaped the MNOs trends globally. However, in order to conclude whether those variations have negatively or positively impacted the MNOs, the revenue trend analysis will be the decisive factor to judge the impact as examined below.

## IV.II  Revenue analysis

As illustrated at figure 4, p. 24, the MNOs have registered total revenue of 1.032 Trillion $ at 2019. The number reflects a great achievement and success of telecommunications industry. In order to deeply explore and analyse the revenue and ARPU trends, the total period of study from 2008 till 2025 have been divided into 3 phases where each period has got different characteristics. As concluded in the major trend's analysis, the MNOs traditional services were increasing till 2012, which presents a successful period for MNOs then the turnover year was 2013 when it started to degrade after. Accordingly, Phase I period has been chosen from 2008 till 2013. Phase II started from 2014 till 2019 which is the present, at this period the OTTs have been evolving fast and has severely reshaped the MNOs trends. Phase III starts from 2020 till 2025 which includes forecasted trends of how MNOs revenues will continue.

**Phase I – Successful Phase for MNOs**

The revenue is increasing steadily at phase I as shown below at Figure 13 (personal collection). The MNOs have been generating highest revenues during this phase from LDC, SMS and voice calls. With this increasing trend, the average revenue growth rate was 12.38 % each year. The ARPU average per year during this phase was 21.7 $ generated from an average subscribers' number of 3.34 billion per year. With these positive indicators, phase I is considered as the reference to evaluate phase II and III. Regression analysis have been done on the available quarterly and annual data of phase I with year number and revenue values as data sets. The analysis results in a positive growing linear trend line with equation of $y=39.836x+729.63$, where y is revenue, and x is the year number. In order to test the accuracy of the derived model formula, the actual revenue of phase I is compared to the revenue derived from the regression analysis as shown at table 8 (personal collection) and Figure 12 (personal collection). The predicted revenue is similar to the actual revenue with an acceptable error rate average of 6.2%. This underlines the accuracy of the derived model and confirms the validity for it to be used to predict the revenues over the years in phase II and Phase III.

| Phase | Year | Actual revenue ($ Billion) | Revenue derived from regression analysis [y=39.836x+729.63] ($ Billion) | Error rate % |
|---|---|---|---|---|
| Phase I | 2008 | 776.84 | 816.88 | 5.2% |
| Phase I | 2009 | 804.27 | 859.25 | 6.8% |
| Phase I | 2010 | 838.85 | 901.63 | 7.5% |
| Phase I | 2011 | 892.97 | 944.00 | 5.7% |
| Phase I | 2012 | 934.94 | 986.37 | 5.5% |
| Phase I | 2013 | 966.47 | 1028.74 | 6.4% |

Table 8. Regression analysis model error rate % (Personal collection)

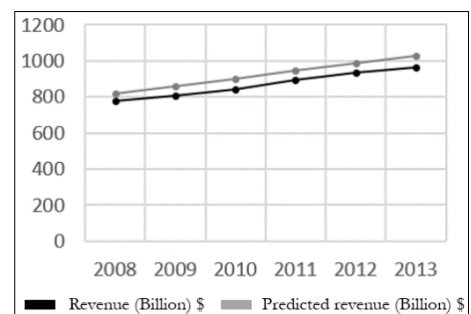

Figure 12. Actual revenues vs predicted revenue (personal collection)

**Phase II – Declining MNOs` Phase with OTT evolution**

As phase I is the success reference for the other two phases, the derived trend from regression analysis of phase I (y=39.836x+729.63) has been used to extrapolate the predicted revenue over phase II and phase III as shown at Figure 13 (personal collection). The actual revenue trend at phase II is increasing, however, as it is obvious from figure 13., the predicted revenue is increasing more rapidly. This means that the revenues after 2013 was increasing, however not with the same growth rate it was increasing in the successful era of phase I. The difference between the actual revenue in phase II and the predicted revenue is considered as the revealed hidden revenue loss in this research which is the grey area in the graph shown at Figure 13. The MNOs were delighted with their actual revenue trend increasing with function of (y=16.365x+878.78), but it shall be increasing with function of (y=39.836x+729.63) referring to phase I as a baseline. The revenue losses numbers of MNOs at phase II is shown below at table 9 (personal collection). MNOs have registered a revenue loss along phase II starting from 2014 till 2019, with an average loss of 175 $ Billion per year. The losses were increasing every year reaching 250 $ billion at 2019.

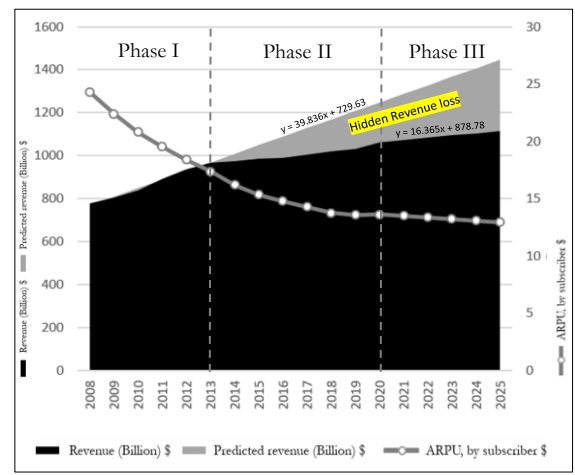

Figure 13. Revenue analysis (personal collection)

| Phase | Year | Actual Revenue (Billion) $ | Predicted revenue derived from regression analysis [y=39.836x+729.63] ($ Billion) | Revenue loss ($ Billion) |
|---|---|---|---|---|
| Phase II | 2014 | 974.88 | 1071.11 | 96.23 |
| Phase II | 2015 | 985.66 | 1113.49 | 127.83 |
| Phase II | 2016 | 991.17 | 1155.86 | 164.69 |
| Phase II | 2017 | 1006.44 | 1198.23 | 191.79 |
| Phase II | 2018 | 1020.79 | 1240.60 | 219.81 |
| Phase II | 2019 | 1032.03 | 1282.97 | 250.94 |

Table 9. Revenue losses of MNOs at Phase II (Personal Collection)

This analysis emphasis the serious issue that MNOs are facing and the negative impact on MNOs that OTTs evolution brought at Phase II specifically.

In order to evaluate the OTTs` trends and further illustrate the revenue shift from MNOs to OTTs. Facebook OTT app has been selected here to evaluate the OTTs performance while MNOs revenue are recording losses. As shown below at Figure 15 (amended from Facebook 2019 and GSMA 2020b), while MNOs ARPU trend is decreasing over the years, Facebook ARPU is rapidly increasing and have crossed MNOs` ARPU at 2016. This means that Facebook generate a revenue from a single user equal to the revenue MNOs generating at 2016. Afterwards, the MNOs ARPU continued to decrease while Facebook ARPU was increasing till it reached 25 $ at 2018 when MNOs` ARPU recorded 14 $.

After concluding the revenue loss of MNOs using regression followed by extrapolation analysis, exact same methodology has been applied on Facebook`s revenue as shown at Figure 14 (personal collection). Surprisingly, while searching a similar revenue loss for Facebook, the results shows a totally opposite case. In Facebook it is not revenue loss, however it is revenue excess. As shown at figure 14 the revenue trend should be increasing with the predicted grey trend; however, the actuality is it is increasing much more than predicted! Table 10 (personal collection) shows the quantified revenue excess that Facebook is generating annually that reached 41 $ Billion at 2018. By looking into table 9 and table 10 together, the revenue excess that Facebook generated every year, is a portion of the revenue losses that MNOs registered. While rest of the MNOs` revenue losses went to other OTTs.

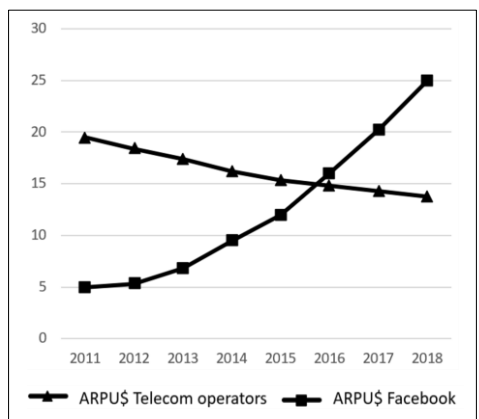
Figure 15. MNOs` vs Facebook ARPU (amended from Facebook 2019 and GSMA Intelligence 2020)

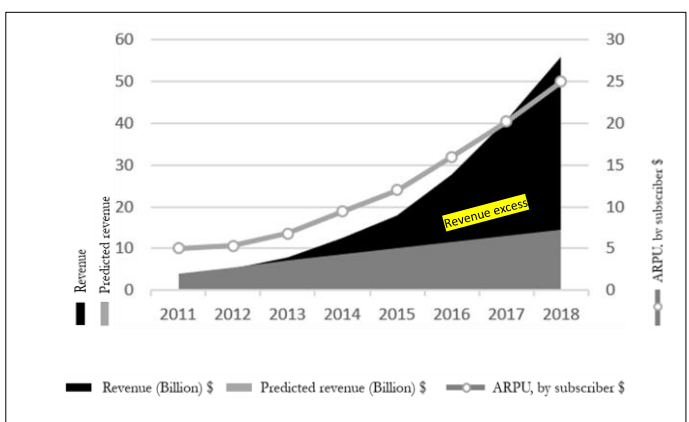
Figure 14. Facebook Revenue vs Estimated revenue growth and ARPU (personal collection)

| Phase | Year | Actual Revenue (Billion) $ | Predicted revenue derived from regression analysis [y=39.836x+729.63] ($ Billion) | Revenue excess ($ Billion) |
|---|---|---|---|---|
| Phase II | 2014 | 12.47 | 8.55 | 3.92 |
| Phase II | 2015 | 17.93 | 10.05 | 7.87 |
| Phase II | 2016 | 27.64 | 11.56 | 16.08 |
| Phase II | 2017 | 40.65 | 13.06 | 27.59 |
| Phase II | 2018 | 55.84 | 14.57 | 41.27 |

Table 10. (Personal collection)

While the MNOs are registering this massive increasing revenues losses year by year. MNOs are under pressure to increase their expenditures in CAPEX and OPEX as shown at Figure 16. (amended from GSMA 2020b) in order to accommodate the data traffic increase caused by OTTs and willing to maintain the customer experience of network`s quality and data speeds. The forced increase in OPEX and CAPEX drives further revenue losses to MNOs unless MNOs follow strategies to decrease expenditures in Phase III.

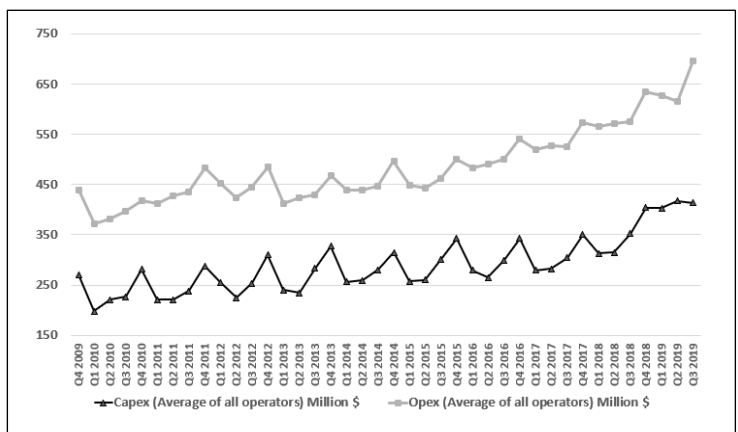
Figure 16. Average CAPEX and OPEX global MNOs (amended from GSMA 2020b)

### Phase III – Opportunity to recover the losses

If the situation remains the same and MNOs will not proactively react against OTTs, the hidden revenue losses will continue to increase as shown at table 11 (personal collection). The MNOs will register a significant loss of 423 $ Billon at 2025. MNOs will be officially a set of pipes that transmitting the OTTs` content, unless the will take serious actions to modernize their business models by implementing proper strategies in order to attract their customers back to their services.

| Phase | Year | Actual Revenue (Billion) $ | Predicted revenue derived from regression analysis [y=39.836x+729.63] ($ Billion) | Revenue loss ($ Billion) |
|---|---|---|---|---|
| Phase III | **2020** | 1060.13 | 1325.35 | 265.21 |
| Phase III | **2021** | 1072.86 | 1367.72 | 294.86 |
| Phase III | **2022** | 1084.79 | 1410.09 | 325.30 |
| Phase III | **2023** | 1094.55 | 1452.46 | 357.91 |
| Phase III | **2024** | 1102.41 | 1494.83 | 392.42 |
| Phase III | **2025** | 1113.45 | 1537.21 | 423.75 |

Table 11. Phase III predicted revenue losses (personal Collection)

As per GSMA (2020a), MNOs planning to invest $1.1 trillion globally in the next five years in their networks. Will OTTs again absorb this investment, or MNOs will follow recommendations to deal with OTTs and modernize their business models?

## V Recommendations

So far, the traditional MNOs, with their response to OTT competition, have been more reactive, rather than proactive. Taking a more proactive stance, ensures a higher and faster return on investment. It is not late for MNOs and telecommunications industry to reclaim lost grounds from OTTs. MNOs have the experience and the right DNA to cope with the situation and turn the table around. Below are major recommendations that MNOs shall follow in order to recover the losses and get their shifted-out customers back.

**Fill the gap:** with the gap analysis done comparing MNOs vs OTTs voice and messaging services features, MNOs shall fill the gap in their offered services and get closer to customers` needs in order to compete with OTTs. Now it is the right time to act, if MNOs are yet to address their customer`s changing demands.

**Quantify the losses and set targets**: Following the hidden loss reveal model developed and examined in this research, MNOs shall apply this model to quantify their losses after defining their success era as a reference. Quantifying the hidden loss reveals the significance of the issue that MNOs encounter, accordingly, helps to set a proper strategy.

**Apply appropriate strategy:** The 3 strategies explored at this research are not necessarily to be mutually exclusive as MNOs should be active in different areas. Categorize your OTTs into 4 groups and 2 dimensions as shown at Figure 17 (personal collection). First dimension is the size of the OTT whether it is big or small based on the usage and popularity in the country where the MNO exist. Second dimension is the overlapping level with the offered services, indirect overlapping and direct based on the offered MNOs` services intersection with OTTs` services. 4 sets of strategies are recommended as shown at figure 17 based on the 2 dimensions:

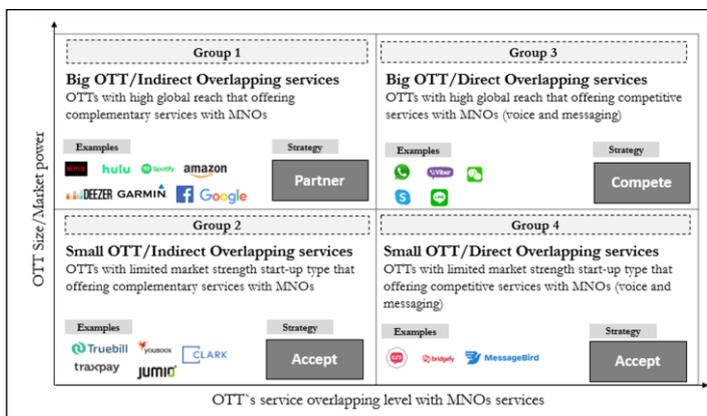

Figure 17. Recommendations (personal collection)

**Group 1:** The OTTs` apps and services which having additive value and may provide access for MNOs to a new target segment of subscribers which can increase their revenues. Such OTTs are firmly established and already have developed a substantial strength in market of music and video. Netflix and Spotify are examples for entertainment type of services in video and music where it can increase MNOs previously limited footprint in these sectors. MNOs have to deal with this group as volume partners to open new market segments or at least to limit their customer churn. For this group the recommended strategy is "Partner".

**Group 2:** The OTTs that offering indirect-overlapping services with MNOs however it is smaller in volume if compared to group 1. Truebill is an example that offers subscriptions tracking services to users that enables them to pay for services using their mobile account. Such OTT can be a value partner to MNOs as it expands their footprint in new areas and represents a value enlargement to MNOs. However, partnering with this group of OTT entails a high flop risk; engaging into partnering in their early development stage before their hype is advantageous for MNOs. For this group the recommended strategy is "Accept".

**Group 3:** While group 1 and 2 of OTTs are not overlapping with MNOs offered services, group 3 has immediate overlap with the offered MNO`s services and entails the major contribution in MNO`s revenues losses. The larger and more competitive the OTT is, the more "Compete" strategy is recommended to be applied by MNOs. As these portions of OTTs such as WhatsApp and Skype are largely overlapping with offered MNOs services of voice, SMS and LDC so it is more appropriate for MNOs to pursue a strategy of "Compete". While four different "compete" strategies discussed at previous section, create own OTT is the most suitable in the current market.

- **Blocking group 3 OTTs is not a recommended option** due to the popularity and the deep penetration of OTTs in the society. As well, OTTs are offering social responsibility services which help the society in different fields.

- **Attack is the best form of defence:** Create own OTTs to compete. With the big customer data that MNOs own, MNOs shall analyse their customer behaviour and develop own OTT app that perfect match their customers and can compete with OTTs. A combined OTT app that includes messaging, voice calls and international calls services with interconnectivity ability to communicate with other existing OTTs, is a suggested OTT.

**Group 4:** OTTs are smaller in size if compared to group 3, their offered services are directly overlapping with MNOs services such as messaging services. They are not presenting high risk on MNOs due to their limited footprint. For such OTTs MNOs shall apply "Accept" strategy. Also, MNOs can acquire such OTTs and develop them to compete with group 3, rather than creating a new OTT app from scratch.

**Reduce OPEX and CAPEX** by Automation and v-RAN, reduction of OPEX will be achieved by network operations automation with a zero human intervention vision. And reduction of Capex will be achieved by modernizing the networks infrastructure to be software centric rather than expressive stagnant hardware-based networks. With the upcoming IOT, 5G and more demanding applications, networks will be extremely complex which will fail to run with the traditional human means to deploy, optimize and operate. It is now mandatory for MNOs to adopt automated learning-based technologies to enable self-driving networks.

## VI Conclusion

With a thorough review of mobile telecommunications trends, variations, and issues during the whole last decade, it is found that telecommunications industry represented in MNOs has reached its peak of success during 2014 then started deteriorating. There are many factors impacted this declining trend, however it has been proven that one of the genuine factors is the evolution of OTTs. From one hand OTTs introduction have led to increase in customers numbers, smartphones and data traffic which forced the MNOs to invest in modernizing their infrastructure to accommodate the increased myriad users' numbers and the tremendous traffic. On the other hand, OTTs have significantly absorbed traditional LDC, voice calls and messaging services from MNOs. The 6 variations in total have led to losses in MNOs` revenues as summarized at figure 18 (personal collection). It has been concluded that the major impacted revenue stream is the SMS and it is predicted to vanish by Q4 2022.

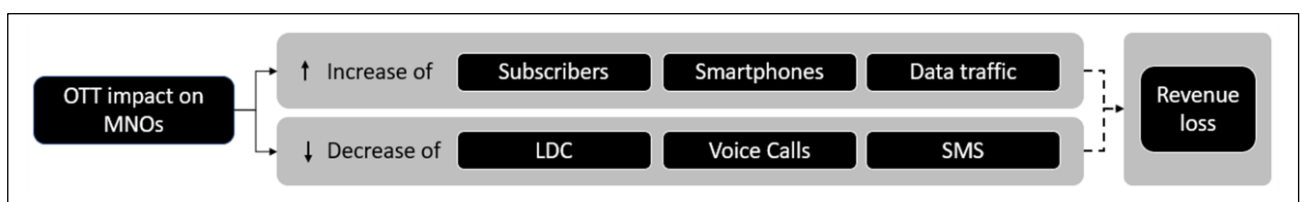

Figure 18. OTTs impact on MNOs (personal collection)

It has been concluded that OTTs have won the battle in the last decade to compete and attract the customers due to their modern offered services which fulfill the customers` needs, unlike the MNOs who have not upgraded their offered limited services` features, this resulted into exponential increase in the usage of OTT services and aggravate the situation of MNOs. OTTs have succeeded in turning the MNOs to reactive technology suppliers rather than proactive. The revenue trend analysis of MNOs during last decade and the correlation with OTT trends proofs that OTTs were threat on MNOs. The less impacted period was from 2008 till 2013 where MNOs revenues were growing steadily however after 2013 the revenue growth started declining. Although the upcoming 5 years trends of MNOs are expected to continue declining, it can be the opportunity to MNOs to recover and get back as the market leaders. Table 12 (personal collection) concludes the major KPIs of global telecommunication industry in 3 phases. This research gives insights for MNOs and regulators to review the severe impact of OTTs on MNOs in a quantified number that shows the significance of this impact.

| KPI | Phase I [2008-2013] | Phase II [2014-2019] | Phase III [2020-2025] | Trend Line |
|---|---|---|---|---|
| Number of subscribers (Billion)(Average) | ↓ 3.3 | → 4.7 | ↑ 5.56 | |
| ARPU $ (Average) | ↑ 21.7 | ↓ 15.7 | ↓ 13.8 | |
| Revenue growth rate (%) (Average) | ↑ 12.3 | ↓ 3.5 | ↓ 1.76 | |
| Revenue sources | Voice calls, SMS, LDC | Data traffic | Data traffic, 5G, IOT | |

Table 12. MNOs` KPIs comparison (personal collection)

It has been concluded that all the previous proposed strategies can be categorized into three holistic approaches of Compete, Partner and Accept. The analysis shows Accept was the most recommended strategy by 54% of the 26 revised researches, followed by 27% Partner and 19% Compete. Unlike other researches that focused on one approach, it was recommend in this research to apply different strategies based on the OTT size and the level of overlapping between MNOs and OTTs offered services. In the background the MNOs must accept the fact that OTTs are using their networks and will continue using. Instead of expanding their network infrastructure only, MNOs must decrease CAPEX and OPEX of their networks by applying the virtualized networks and the automation that can save up to 30% of their expenditures. The reason behind promoting Accept strategy is that the proliferation of OTTs is not stopping here, however it is concluded that the 5G will empower the current OTTs with better speeds and low latency and will generate a new wave of OTTs in the near future.

The OTT impact on MNOs subscribers have been analyzed in this research, however with the introduction of IOT, the MNOs will provide connections to the machines including cars, hospitals, drones, etc. Who will monetize the connections of the machines? is it the MNOs who provide the connectivity, or the new wave of the OTTs will ride into the scene again? The new wave of OTTs and their impact on MNOs is a topic yet to be explored in researches.

## GLOSSARY OF ACRONYMS

All the acronyms used in this dissertation are listed below for readers convenience

| | |
|---|---|
| ARPU | Average Revenue Per User |
| CS | Circuit Switching |
| CAGR | Compound Annual Growth Rate |
| GDP | Gross Domestic Product |
| IMS | IP Multimedia Subsystem |
| IOT | Internet Of Things |
| IP | Internet Packets |
| KPI | Key Performance Indicator |
| LTE | Long Term Evolution |
| MNO | Mobile Network Operator |
| OTT | Over The Top |
| PS | Packet Switching |
| QoS | Quality of Service |
| SMS | Short Message Service |
| SWOT | Strengths Weaknesses Opportunities Threats |
| UN | United Nations |
| USA | United States of America |
| VoIP | Voice over Internet Protocol |
| VOLTE | Voice Over LTE |
| 3G | 3rd Generation mobile network |
| 4G | 4th Generation mobile network |
| 5G | 5th Generation mobile network |

**ACKNOWLEDGMENT**

I would like to thank all my supervisors specially who have given me guidance and constructive feedbacks throughout this project and during all this master's degree journy. Without their guidance and persistent help this master's degree and dissertation would not have been achievable. A special thanks to the influencers in telecommunications industry who have inspired me to complete this research and kept me inquisitive to help improving this industry.